\documentclass[12pt]{iopart} 

\usepackage{times}

\begin{document} 

\title[]{The Issue of Time in Quantum Geometrodynamics} 

\author{Nathan D. George\dag, Adrian~P.~Gentle\dag, Arkady~Kheyfets\ddag and 
        Warner~A.~Miller\dag\S}
\address{\dag\ Theoretical Division (T-6, MS B227) Los Alamos National
         Laboratory, Los Alamos, NM 87545, USA}
\address{\ddag\ Department of Mathematics, North Carolina State University,
         Raleigh, NC 27695-8205, USA.}
\address{\S\ Department of Physics, Florida Atlantic University, Boca Raton, 
         FL 33431, USA}

\begin{abstract} 
  
  Standard techniques of canonical gravity quantization on the
  superspace of 3--metrics are known to cause insurmountable
  difficulties in the description of time evolution. We forward a new
  quantization procedure on the superspace of true dynamic variables
  -- geometrodynamic quantization.  This procedure takes into account
  the states that are ``off-shell'' with respect to the constraints
  and thus circumvents the notorious problems of time.  In this
  approach quantum geometrodynamics, general covariance, and the
  interpretation of time emerge together as parts of the solution to
  the total problem of geometrodynamic evolution.

\end{abstract}

\ead{\mailto{ndg@lanl.gov}, \mailto{apg@lanl.gov}, 
     \mailto{kheyfets@math.ncsu.edu}, \mailto{wam@fau.edu}}

\bigskip 

The standard approach to canonical quantum gravity \cite{Isham99,
  Kuc93} is based on the classical dynamic picture of the evolving
3--geometry of a slicing of a spacetime manifold described by the
lapse function $N$ and the shift functions $N^i$. The canonical
variables are the 3--metric components $g_{ik}$ on a spatial slice
$\Sigma$ of the foliation induced by the spacetime 4--metric, and
their canonical conjugate momenta $\pi^{ik}$.  The customary
variational procedure applied to the Hilbert action expressed in terms
of these canonical variables yields Hamilton dynamics that, after
applying the canonical quantization procedure on the superspace of
3--metrics (in both Dirac's and ADM square root Hamiltonian
approaches), produces a quantum theory that appears to be incapable of
providing a consistent description of time evolution for quantum
gravitational systems. The source of the difficulties can be traced to
mixing dynamical considerations with the requirements of general
covariance and to restricting quantum states to the shell determined
by constraints.

The situation changes dramatically if York's analysis of gravitational
degrees of freedom \cite{Yor72} is taken into account and actively
utilized.  According to York, the set of six parameters describing the
slice 3--metric should be split into two subsets, $\{\beta_1,
\beta_2\}$ (two functions) and $\{\alpha_1, \alpha_2, \alpha_3,
\Omega\}$. The first of these is treated as the set of true
gravitational degrees of freedom (the initial values for them can be
given freely), while the second is considered to be the set of
embedding variables. The $\alpha$ parameters are often referred to as
coordinatization parameters, while $\Omega$ is called, depending on
the context, the slicing parameter, the scale factor, or the
many--fingered time parameter. Information relevant to dynamics is
carried by $\beta$ parameters, while $\alpha$ and $\Omega$ essentially
describe time. The true dynamic variables form what we call a dynamic
superspace while the embedding variables are treated as functional
parameters.

The idea is to develop geometrodynamics from the very beginning on the
dynamic superspace instead of the superspace of 3--metrics or
3--geometries.  The variational principle on the dynamic superspace or
its phase space (formed by true dynamic variables $\{\beta_1,
\beta_2\}$ and their conjugate momenta $\{\pi_{\beta_1},
\pi_{\beta_2}\}$) yields the dynamic equations describing the
evolution of the true dynamic variables. All of these equations depend
on lapse and shift and contain embedding variables as functional
parameters. These are treated as an external field and are determined
by additional equations that do not follow from the variational
principle on dynamic superspace. The quantization procedure is
performed on the dynamic superspace (only ${\beta}$-s are quantized,
i.~e.~generate commutation relations, while the embedding variables
form a classical field). The Schr\" odinger equation is obtained by a
quantization procedure from the Hamilton--Jacobi equation on the
dynamic superspace and describes the time evolution of the state
functional on the true dynamic superspace coupled with the external
classical field determined by the embedding variables. Such a coupling
can be achieved via a procedure similar to that of Hartree--Fock.

In a more detailed and precise description that follows, we omit
indices on variables $\beta$ and $\alpha$ for the sake of notational
simplicity. They can be recovered easily whenever necessary.

We start from the standard Lagrangian ${\cal L}$ (written in terms of
the 3--metric, shift and lapse) and the associated action (with
appropriate boundary terms, as needed, to remove the second time
derivatives terms) and we introduce the momenta conjugate to the true
dynamic variables
\begin{equation} 
\label{301} 
\pi_\beta = {\partial {\cal L}\over \partial\dot\beta}. 
\end{equation} 
We then use these $\pi_\beta$'s to form the geometrodynamic Hamiltonian 
${\cal H}_{dyn}$, 
\begin{equation} 
\label{302} 
{\cal H}_{dyn} = \pi_\beta \dot\beta - {\cal L}. 
\end{equation} 
The arguments of the Hamiltonian ${\cal H}_{dyn}$ are described by 
the expression
\begin{equation} 
\label{303} 
{\cal H}_{dyn} = {\cal H}_{dyn}(\beta , \pi_\beta  ; \Omega ,  
\alpha ). 
\end{equation}
The variables following the semicolon are treated as describing an
external field, while the ones preceding the semicolon are the
coordinates and momenta of the true gravitational degrees of freedom,
i.e. of the true geometrodynamics. The variation of $\beta$ and
$\pi_\beta$ produce Hamilton equations on the dynamic superspace,
while variation of the ends leads to the Hamilton--Jacobi equation
\begin{equation} 
\label{306}  
{\delta S\over\delta t} = - {\cal H}_{DYN}\left(\beta , 
{\delta S\over\delta\beta}; \Omega ,  
\alpha \right). 
\end{equation} 
Here $S$ is a functional of $\beta$ and, in addition, a function of $t$, 
\begin{equation} 
\label{307} 
S = S\left[\beta ; t \right) . 
\end{equation} 
and ${\delta \over \delta t}$ is defined by  
\begin{equation} 
\label{307a} 
{\partial S \over \partial t} = \int {\delta S\over\delta t} d^3x. 
\end{equation} 

The Hamilton--Jacobi equation (\ref{306}) is incapable of providing
any predictions as its solutions depend on the functional parameters
$\Omega$ and $\alpha$ which are not yet known.  One can complete the
picture by adding the standard constraint equations of general
relativity, obtained by variations of shift and lapse.  These
constraints should be satisfied once the solution for the
geometrodynamic variables $\beta$, $\pi_\beta$ (with appropriate
initial data) is obtained and substituted. Using the symbols $[\beta
]_s$, $[\pi_\beta ]_s$ for such a solution, we have
\begin{eqnarray} 
\label{308} 
{\cal H}^i\left( [\beta ]_s, [\pi_\beta ]_s , 
  \Omega , \alpha\right) & = & 0, \\
{\cal H}\left( [\beta ]_s, [\pi_\beta ]_s ,  
\Omega , \alpha\right) & = & 0. 
\end{eqnarray} 
These constraint equations should be treated as additional symmetries,
or the equations for an external field. They do follow from the shift
and lapse invariance of the action but their derivation in this new
setting depends on the structure of the whole action integral.  As a
result, they cannot replace the full set of equations for
geometrodynamic evolution (which is usually done on the superspace of
3--metrics). However, the resulting complete system of equations is
equivalent to that of the standard geometrodynamics on the superspace
of 3--geometries \cite{KheMil96}.

For the purpose of quantization, we make a transition to the
corresponding Schr\"odinger equation based entirely on dynamics and
ignoring the system symmetries
\begin{equation} 
\label{309} 
i \hbar {\delta\Psi\over\delta t} = \widehat{\cal H}_{dyn}\left(\beta , 
\widehat\pi_\beta ; \Omega , \alpha\right) \Psi 
\qquad \mbox{where} \qquad \widehat\pi_\beta = {\hbar\over i} {\delta\over\delta\beta}.
\end{equation} 
The Schr\"odinger equation (\ref{309}) implies that commutation
relations are imposed only on the true dynamic variables and treats
the embedding variables as external classical fields. The state
functional $\Psi$ in this equation is a functional of $\beta$ and a
function of $t$,
\begin{equation} 
\label{310} 
\Psi = \Psi\left[\beta , t\right) .
\end{equation} 
This Schr\"odinger equation (with specific initial data) can be solved
(cf., for instance the example of the Bianchi~1A cosmological model
\cite{KheMil94a,KheMil96}). The resulting solution $\Psi_s$ of this
Schr\"odinger equation is not capable of providing any definite
predictions as it depends on four functional parameters $\Omega$,
$\alpha$ which remain at this stage undetermined. All expectations,
such as the expectation values of $\beta$
\begin{equation} 
\label{311} 
<\beta >_s = \langle\Psi_s\vert\beta\vert\Psi_s\rangle = 
\int\Psi^*_s \beta \Psi_s \, {\cal D}\beta 
\end{equation}
or of $\widehat\pi_\beta$ 
\begin{equation} 
\label{312} 
<\pi_\beta >_s = \langle\Psi_s\vert\widehat\pi_\beta\vert\Psi_s\rangle = 
\int\Psi^*_s \widehat\pi_\beta \Psi_s \, {\cal D}\beta 
\end{equation} 
also depend on these functional parameters.  To specify these
functions we resort to the constraint equations. The treatment of the
constraints has nothing to do with the quantization of
geometrodynamics.  It merely introduces the coupling between the
already quantized geometrodynamics and the classical field determined
by the embedding variables.  In other words, the constraints take care
of the symmetries which are classical in nature to the extent that
they are capable of doing so.

As in case of classical geometrodynamics, we impose the constraints on
the solution of the dynamic equations (Schr\"odinger equation) with
appropriate initial data and in this way, determine the unique values
of $\Omega$ and $\alpha$. It is possible that there are several ways
to couple the constraints to the quantization of the true dynamic
variables, $\beta$. Here we impose the four constraints only on the
expectation values of the conformal dynamics
\begin{equation} 
\label{313} 
\matrix{{\cal H}^i\left( <\beta >_s, <\pi_\beta >_s , 
\Omega , \alpha\right) = 0 \cr 
  \cr 
{\cal H}\left( <\beta >_s, <\pi_\beta >_s ,  
\Omega , \alpha\right) = 0. \cr}
\end{equation}  
Lapse and shift are assumed to be given either explicitly or by
additional conditions.

Evolution can be described as follows. Initial data at $t = t_0$
consist of the initial state functional $\Psi = \Psi_0$ and the
initial values (functions) of embedding variables. In addition, lapse
and shift are supposed to be given either explicitly or by additional
conditions. Equations (\ref{311}), (\ref{312}) yield the expectation
values (functions) of the true dynamic variables and their conjugate
momenta. The results are substituted into the constraints (\ref{313}).
After this, the constraints are solved with respect to the time
derivatives of embedding variables. A step forward in time (say, with
the increment $\Delta t$) is performed by integration of the
constraints to evolve the embedding variables and by integration of
the Schr\" odinger equation (\ref{309}) to evolve the state
functional.  This concludes one step forward in time. The next step is
performed by repeating the same operations in the same order.

One can be referred to \cite{KheMil94a,KheMil96} for two particular
examples illustrating such geometrodynamic evolution for the Bianchi
1A and Taub cosmologies respectively. The first one can and has been
solved analytically, while the latter one has been solved numerically.

The resulting canonical gravity quantization procedure circumvents all
the standard problems of time and removes all the obstacles for
describing the time evolution of quantum gravitational systems. This
has been achieved by including ``off--shell'' quantum states and
imposing the constraints only on the expectation values of the dynamic
variables.

It should be stressed that all three components of the evolution
procedure described for quantum geometrodynamic systems --- quantum
dynamics itself, constraints enforcing the symmetries (general
covariance), and the interpretation of time --- emerge together as the
solution to the total problem of geometrodynamic evolution.

\end{document}